\documentclass[11pt]{article}
\pdfoutput=1

\usepackage{a4wide}
\usepackage[fleqn]{amsmath}

\RequirePackage{amsmath,amssymb,amsxtra,amsthm}

\RequirePackage[T1]{fontenc}
\RequirePackage[utf8]{inputenc}

\RequirePackage{mathrsfs}
\RequirePackage{mathpazo}

\RequirePackage{setspace}
\RequirePackage{array}
\RequirePackage{booktabs}

\RequirePackage{braket}

\RequirePackage[pdftex,final]{graphicx}
\graphicspath{{Plots/}}

\usepackage{cite}

\newcommand{\be}{\begin{equation}}
\newcommand{\ee}{\end{equation}}
\newcommand{\bea}{\begin{eqnarray}}
\newcommand{\eea}{\end{eqnarray}}

\newcommand{\cF}{\mathcal{F}}

\numberwithin{equation}{section}
\numberwithin{table}{section}
\numberwithin{figure}{section}

\author{
  \begin{minipage}{0.97\linewidth}
    \vspace{1cm}
    \begin{center}
      \begin{small}
        \textbf{Carlo Angelantonj}$^{1}$, \textbf{Ioannis Florakis}$^{2}$ and  \textbf{Mirian Tsulaia}$^{3}$
     \end{small}
    \end{center}
    \vspace{.3cm} \hspace{1.3cm}\begin{minipage}{.75\linewidth}
      {\it \begin{footnotesize}
          \begin{itemize}
          \item[${}^1$] Dipartimento di Fisica, Universit\`a di Torino, and INFN Sezione di Torino
          \\
            Via P. Giuria 1, 10125 Torino, Italy
          \item[${}^2$] Max-Planck-Institut f\"{u}r Physik,\\
	    Werner-Heisenberg-Institut,
          80805 M\"{u}nchen, Germany
          \item[${}^3$] Faculty of Education Science Technology and Mathematics, \\
          University of Canberra, Bruce ACT 2617, Australia
          \end{itemize}
        \end{footnotesize}}
    \end{minipage}
    \vspace{1cm}
  \end{minipage}
}

\date{}

\title{\vspace{3cm}
  \begin{huge}
    \textbf{Universality of Gauge Thresholds \\[10pt] in Non-Supersymmetric Heterotic Vacua }
  \end{huge}
}

\begin{document}

\begin{titlepage}
  \maketitle
  \thispagestyle{empty}

  \vspace{-14cm}
  \begin{flushright}
    MPP-2014-315
   \end{flushright}

  \vspace{11cm}

  \begin{center}
    \textsc{Abstract}\\
  \end{center}

We compute one-loop threshold corrections to non-abelian gauge couplings in four-di\-men\-sional heterotic vacua with spontaneously broken $\mathcal{N} = 2 \to \mathcal{N}=0$ supersymmetry, obtained as Scherk-Schwarz reductions of six-dimensional K3 compactifications. As expected, the gauge thresholds are no-longer BPS protected, and receive  contributions also from the excitations of the RNS sector. Remarkably, the difference of  thresholds for non-abelian gauge couplings is BPS saturated and exhibits a universal behaviour independently of the orbifold realisation of K3. Moreover, the thresholds and their difference develop infra-red logarithmic singularities whenever charged BPS-like states,
originating from the twisted RNS sector, become massless at special loci in the classical moduli space.

\vfill
{\small
\begin{itemize}
\item[E-mail:] {\tt carlo.angelantonj@unito.it}\\ {\tt florakis@mppmu.mpg.de}\\
{\tt mirian.tsulaia@canberra.edu.au}
\end{itemize}
}
\vfill

\end{titlepage}

\setstretch{1.1}



\section{Introduction}

In the last decades we have witnessed a tremendous progress in  understanding the structure of supersymmetric vacua in String Theory and M/F-theory. Several semi-realistic vacua that incorporate the salient features of the MSSM have been constructed and analysed to a remarkable extent. Their low-energy effective action with $\mathcal{N}=1$ supersymmetry has been fully reconstructed at tree-level, and the incorporation of quantum and $\alpha '$ corrections is still a subject of intense study.  Despite these successful endeavours, supersymmetry breaking in String Theory remains a compelling open problem that string phenomenology aspires to address. 

A fully-fledged approach to spontaneous supersymmetry breaking in String Theory, that admits an exactly solvable world-sheet description, is the stringy realisation \cite{SSstringi,SSstringii,FKPZ,SSstringiii} of the Scherk-Schwarz mechanism \cite{Scherk:1978ta,Scherk:1979zr}, via special freely-acting orbifolds. In this class of vacua, the supersymmetry breaking scale is tied to the size of compact dimensions, while the exponential growth of string states may destabilise the classical vacuum due to the emergence of tachyonic excitations. This is closely related to the Hagedorn problem of String Thermodynamics \cite{Atick:1988si} and can be circumvented in special constructions \cite{Angelantonj:2006ut, Angelantonj:2008fz, Florakis:2010ty}. Moreover,  it has been recently argued that closed string tachyons emerging from twisted orbifold sectors of a class of heterotic vacua with explicitly broken supersymmetry can actually acquire a mass by blowing-up the orbifold singularities \cite{Nibbelink:2014ula}. 

In all those cases where supersymmetry is (spontaneously) broken but the vacuum is classically stable, it is meaningful and important to study one-loop radiative corrections to couplings in the low-energy effective action.  The emergence of one-loop tadpoles for massless states does not impinge on the validity of the one-loop analysis, although it makes the incorporation of higher loops problematic, unless the back-reaction on the classical vacuum is properly taken into account \cite{Fischler:1986ci, Fischler:1986tb}. 

For this reason, we address in this letter the problem of computing one-loop threshold corrections to gauge couplings in a class of four-dimensional heterotic vacua with spontaneously broken supersymmetry, that can be built as ${\rm K3}$ reductions of the  ${\rm SO} (16) \times {\rm SO} (16)$ construction of \cite{Itoyama:1986ei} in terms of freely-acting orbifolds. In contrast to heterotic vacua with unbroken supersymmetry, where the moduli dependence of the one-loop corrected gauge couplings arises from the BPS sector, in the case of
 spontaneously broken supersymmetry the amplitude receives contributions from the full tower of charged string states, and is no-longer topological. Nevertheless, we find 
that the difference between gauge thresholds exhibits a remarkable {\em universal} structure akin to the $\mathcal{N}=2$ supersymmetric case, due to highly non-trivial cancellations induced by an MSDS spectral flow \cite{Kounnas:2008ft,Florakis:2009sm,Faraggi:2011aw} in the bosonic right-moving sector of the heterotic string. 

The striking signature of spontaneous supersymmetry breaking is the emergence of logarithmic singularities at special points of the classical moduli space. These are ascribed to charged BPS-like states that become massless at points of gauge symmetry enhancement, and survive in the difference of gauge thresholds.

The paper is structured as follows: in Section 1 we define the freely-acting orbifold  responsible for the spontaneous breaking of supersymmetry and present the corresponding  one-loop partition function. Section 2 is devoted to the evaluation of gauge threshold corrections for the non-abelian gauge couplings and contains the main results of our investigation. Finally, in Section 3 we discuss the relevant decompactification limits and comment on their physical interpretation.

\section{Heterotic vacuum with spontaneous supersymmetry breaking}

The class of non-supersymmetric  vacua that we shall focus on is obtained as a Scherk-Schwarz reduction of six-dimensional K3 compactifications of the ${\rm E}_8 \times {\rm E}_8$ heterotic string. They can also be viewed as K3 reductions of the Itoyama-Taylor vacuum \cite{Itoyama:1986ei}, that corresponds to a lower-dimensional freely-acting implementation  of the  non-supersymmetric, non-tachyonic, ${\rm SO} (16) \times {\rm SO} (16)$ construction  \cite{AlvarezGaume:1986jb, Dixon:1986iz}.

For concreteness, we shall consider the $T^6/\mathbb{Z}_N \times \mathbb{Z}_2^\prime$ compactification of the ten-dimen\-sio\-nal ${\rm E}_8 \times {\rm E}_8$ heterotic string, with factorised $T^6 = T^4 \times T^2$. The $\mathbb{Z}_N$, with $N=2,3,4,6$ rotates chrystallographycally the complexified $T^4$ coordinates as
\begin{equation}
v\, :\quad z_1 \to e^{2i\pi/N}\, z_1\,, \qquad z_2\to e^{-2i \pi/N} \, z_2\,,
\end{equation}
and realises the singular limit of the K3 surface, preserving 8 supercharges. The $\mathbb{Z}_2^\prime$ is instead freely acting and is generated by
\begin{equation}
v' = (-1)^{F_{\rm st} + F_1 + F_2}\, \delta\,.
\end{equation}
Here, $F_{\rm st}$ is the space-time fermion number,  responsible for the breaking of supersymmetry, $F_1$ and $F_2$ are the ``fermion numbers'' of the two original $E_8$'s, whereas $\delta$ acts as an order-two shift along the remaining $T^2$. The combined action of $\delta$ and $(-1)^{F_{\rm st}}$ is responsible for the spontaneous breaking of the $\mathcal{N} = 2$ supersymmetry down to $\mathcal{N} =0$, while the presence of $(-1)^{F_1 + F_2}$ guarantees the classical stability of the vacuum\footnote{This is no-longer true when Wilson lines are turned on, whereby all non-supersymmetric heterotic vacua can be continuously connected \cite{GinspargWR,NairZN}. In this note we shall always assume a trivial Wilson-line background.}. 

The one-loop partition function reads
\begin{equation}
\begin{split}
\mathcal{Z}=&\tfrac{1}{2} \sum_{H,G=0}^1\tfrac{1}{N} \sum_{h,g=0}^{N-1}\left[\tfrac{1}{2}\sum_{a,b=0}^1 (-)^{a+b}
\vartheta \left[^{a/2}_{b/2}\right]^2 \,
\vartheta \left[^{a/2+h/N}_{b/2+g/N} \right]\,
\vartheta \left[^{a/2-h/N}_{b/2-g/N}\right]
\right]
\\
&\qquad\times \left[\tfrac{1}{2}\sum_{k,\ell=0}^1
\bar\vartheta \left[^{k/2}_{\ell/2}\right]^6\,
\bar\vartheta \left[^{k/2+h/N}_{\ell/2+g/N} \right]\,
\bar\vartheta \left[^{k/2-h/N}_{\ell/2-g/N} \right]
\right]\,
\left[\tfrac{1}{2}\sum_{r,s=0}^1\bar\vartheta \left[^{r/2}_{s/2} \right]^8\right] 
\\
&\qquad\times \,\frac{1}{\eta^{12}\bar\eta^{24}}
\,(-)^{H(b+\ell+s)+G(a+k+r)+HG}\, \varGamma_{2,2} \left[^H_G \right] 
\,\varLambda^{\rm K3} \left[^h_g\right] \,.
\end{split}\label{partition}
\end{equation}
Here, $\eta$ is the Dedekind function and $\vartheta \left[{\alpha \atop\beta} \right]$ are the standard Jacobi theta constants with characteristics. The sum over of the spin structures $a$, $b$, $k$, $\ell$, $r$ and $s$ yield the 
ten-dimensional ${\rm E}_8 \times {\rm E_8}$ heterotic-string spectrum, while $(h,g)$ and $(H,G)$ correspond to the $\mathbb{Z}_N$ and $\mathbb{Z}_2^\prime$ orbifolds.
The two-dimensional Narain lattice with characteristics is defined as
\begin{equation}
\varGamma_{2,2} \left[^H_G\right]= \tau_2 \sum_{\vec m , \vec n} e^{i \pi G (\vec \lambda_1 \cdot \vec m + \vec \lambda_2 \cdot \vec n )} \, \varGamma_{\vec m + \frac{H}{2} \vec\lambda_2, \vec n + \frac{H}{2} \vec \lambda_1} (T,U)\,,
\end{equation}
with
\begin{equation}
\varGamma_{\vec m , \vec n} (T,U) = q^{\frac{1}{4 T_2 U_2 } |m_2 - U m_1 + \bar T \, (n^1 + U n^2)|^2}\, \bar q^{\frac{1}{4T_2 U_2 } |m_2 - U m_1 +  T \, (n^1 + U n^2)|^2}
\,,
\end{equation}
and depends on the  K\"ahler and complex structure moduli $T$ and $U$. 
As usual, momenta and windings are labeled by $\vec m$ and $\vec n$, while the integral vectors $\vec \lambda_1$ and $\vec \lambda_2$ encode the freely-acting shift of $\mathbb{Z}_2^\prime$. Without loss of generality, we shall focus on the case $\vec\lambda_1 = (1,0)$ and $\vec \lambda_2 = (0,0)$ corresponding to a momentum shift along the first $T^2$ direction. All other cases can be related to the former by suitable redefinitions of the $T$ and $U$ moduli.

Finally,
\begin{equation}
\varLambda^{\rm K3} \left[^h_g \right] = \left\{
\begin{array}{cl}
\varGamma_{4,4} &\qquad {\rm for}\ (h,g)=(0,0)\,,
\\[10pt]
 \frac{ k \left[^h_g \right]\, |\eta |^{12}}{ \left| \vartheta \left[ {1/2+h/N \atop 1/2 +g/N}\right]\, 
\vartheta \left[ {1/2-h/N \atop 1/2 - g/N}\right] \right|^2} &\qquad {\rm for}\ (h,g)\not= (0,0)\,,
\end{array}
\right.
\end{equation}
with $\varGamma_{4,4}$ being the conventional Narain lattice associated to the $T^4$,  $k \left[^0_g \right] = 16\, \sin^4 (\pi g/N)$ counting the number of twisted sectors of the $\mathbb{Z}_N$ orbifold, and the remaining $k \left[^h_g \right]$'s with $h\neq 0$ being determined by modular invariance.

As a consequence of the Scherk-Schwarz mechanism, the two gravitini acquire a mass  $m_{3/2}  = |U| /\sqrt{T_2\, U_2}$ , and supersymmetry is spontaneously broken at a generic point in the classical moduli space. The $\mathbb{Z}_2^\prime$ also breaks the ${\rm E}_8 \times {\rm E}_7$ gauge group of the $\mathcal{N}=2$ theory down to ${\rm SO} (16) \times {\rm SO} (12)$, up to abelian factors. The full spectrum can be derived from \eqref{partition} using standard techniques.

Notice that, as in the parent ten-dimensional ${\rm SO} (16) \times {\rm SO } (16)$ non-supersymmetric theory \cite{AlvarezGaume:1986jb, Dixon:1986iz}, the spectrum is free of tachyonic excitations at a generic point of the $(T,U)$ moduli space. This can be verified by looking at the $H\neq 0$, $a= 0$ contributions to \eqref{partition}.

\section{One-loop thresholds for non-abelian gauge couplings}

Although the vacuum configuration presented in the previous section is not supersymmetric, the absence of physical tachyons in the perturbative spectrum implies that it is classically stable. As a result, it is fully justified and important to study one-loop radiative corrections to couplings in the low-energy effective action, in contrast to higher-loop diagrams that diverge due to the emergence of one-loop tadpoles  back-reacting on the vacuum \cite{Fischler:1986ci,Fischler:1986tb}. This is still an open problem in String Theory, and has recently triggered a growing interest \cite{Dudas:2004nd,Kitazawa:2008hv,Pius:2014gza}.

To this end, we shall address here the question of quantum corrections to the couplings of the non-abelian ${\rm SO} (16) \times {\rm SO} (12)$ gauge factors, extending the analysis of \cite{Dixon:1990pc} to non-supersymmetric vacua. 

Threshold corrections $\varDelta_\mathcal{G}$ associated to the group factor $\mathcal{G}$ appear in the relation between the running gauge coupling $g_\mathcal{G}^2 (\mu )$ of the low-energy theory and the string coupling $g_{\rm s}$
\begin{equation}
\frac{16 \pi^2}{g_\mathcal{G}^2 (\mu)} = \frac{16\pi^2}{g_{\rm s}^2} + \beta_\mathcal{G} \, \log \frac{M_{\rm s}^2}{\mu^2} + \varDelta_\mathcal{G}\,,
\end{equation}
where, in the case at hand, the Kac-Moody algebra is realised at level one, and $M_{\rm s}$ sets the string scale. They encode the contribution of the infinite tower of massive string states to the one-loop diagram, and can be organised as
\begin{equation}
\begin{split}
\varDelta_\mathcal{G} &\equiv \frac{i}{2 \pi \, N}\, {\rm R.N.} \int_\mathcal{F} d\mu\, 
\sum_{H,G=0}^1\, \sum_{h,g=0}^{N-1}\, \varDelta_{\mathcal{G}} \left[ ^{H,\ h}_{G,\ g}\right]
\\
&= \frac{i}{2 \pi \, N}\, {\rm R.N.} \int_\mathcal{F} d\mu\, 
\sum_{H,G=0}^1\, \sum_{h,g=0}^{N-1}\, (-1)^{HG}\,
\frac{\mathcal{L} \left[^{H,\ h}_{G,\ g}\right]}{\eta^2}\,
\frac{\varPhi_{\mathcal{G}} \left[^{H,\ h}_{G,\ g}\right]}{\bar\eta^{18}}  \,
\frac{\varLambda^{{\rm K3}} \left[^h_g \right]}{\eta^4 \, \bar\eta^4} \,
\frac{\varGamma_{2,2} \left[^H_G \right] }{\eta^2 \, \bar \eta^2}
\,.
\end{split}
\end{equation}
In this expression, $d\mu$ denotes the ${\rm SL} (2; \mathbb{R})$ invariant measure, while ${\rm R.N.}$ stands for the modular-invariant prescription of \cite{Angelantonj:2011br, Angelantonj:2012gw} for regularising the infra-red divergences of the integral. 

The quantity $\mathcal{L} \left[^{H,\ h}_{G,\ g} \right]$  encodes the spin-structure sum over the integrated world-sheet
correlators for the four-dimensional space-time fields, whereas  $\varPhi_{\mathcal{G}}\left[^{H,\ h}_{G,\ g}\right]$  encodes the contribution of the gauge sector with the relevant trace insertion. They are defined as
\begin{equation}
\mathcal{L} \left[^{H,\ h}_{G,\ g}\right] \equiv \tfrac{1}{2}\sum_{(a,b)\neq(1,1)}(-)^{a(1+G)+b(1+H)}\, \partial_\tau\left(\frac{\vartheta \left[^{a/2}_{b/2}\right]}{\eta}\right)\,
\frac{\vartheta \left[^{a/2}_{b/2}\right] \, \vartheta \left[^{a/2+h/N}_{b/2+g/N}\right]\, \vartheta \left[^{a/2-h/N}_{b/2-g/N}\right]}{\eta^3} \,,
\end{equation}
and
\begin{equation}
\begin{split}
\varPhi_{\mathcal G} \left[^{H,\ h}_{G,\ g}\right] &\equiv \tfrac{1}{4} \left(\frac{1}{(2\pi i)^2}\partial_{z_{\mathcal{G}}}^2-\frac{1}{4\pi\tau_2}\right)
\Biggl[\sum_{k,\ell =0}^1 (-)^{kG+\ell H}\,\bar\vartheta \left[^{k/2}_{\ell /2} \right]^6\,\bar\vartheta \left[^{k/2+h/N}_{\ell /2+g/N}\right]\,\bar\vartheta \left[^{k/2-h/N}_{\ell/2-g/N} \right]
\\
& \qquad \times \sum_{r,s=0}^1(-)^{r G+s H}\,\bar\vartheta \left[^{r/2}_{s/2}\right]^8\Biggr](z_{\mathcal{G}})\Bigr|_{z_\mathcal{G}=0}\,.
\end{split}
\end{equation}
In the latter equation, it is implied that the VEV $z_{\mathcal G}$ is only inserted along the particular theta function corresponding to the Cartan charge whose group trace we are considering. 

It is convenient to arrange the $4  N^2$ sectors of the orbifold so as to distinguish the origin of the various contributions to the thresholds. The $(h,g)=(0,0)$ sector corresponds to the Itoyama-Taylor construction \cite{Itoyama:1986ei} reduced to four-dimensions, and is
proportional to the $T^4$ lattice $\varLambda^{\rm K3} \left[ {0\atop 0}\right]$, depending on the invariant $T^4$ moduli. Furthermore, since $h=g=0$,   one is effectively dealing with the  ${\rm SO} (16) \times {\rm SO} (16)$ lattice. Hence, the group traces  are independent of the choice of gauge group $\mathcal G$, implying that the difference of thresholds is independent of the $T^4$ moduli.

An explicit calculation yields
\begin{equation}
\begin{split}
 \varDelta_{\varLambda} &=-\frac{1}{2N\times 12^2}\,\frac{\varLambda^{\rm K3} \left[^0_0 \right]}{\eta^{12}\,\bar\eta^{24}} \left[ \varGamma_{2,2} \left[^0_1 \right]~(\vartheta_3^8-\vartheta_4^8)~\bar\vartheta_3^4\,\bar\vartheta_4^4\,\left( (\hat{\bar E}_2-\bar\vartheta_3^4)\,\bar\vartheta_3^4\,\bar\vartheta_4^4+8\bar\eta^{12}\right) \right. 
 \\
&\qquad -\varGamma_{2,2} \left[^1_0\right] ~(\vartheta_3^8-\vartheta_2^8)~\bar\vartheta_3^4\,\bar\vartheta_2^4\,\left( (\hat{\bar E}_2+\bar\vartheta_3^4)\,\bar\vartheta_2^4\,\bar\vartheta_3^4-8\bar\eta^{12}\right)
\\
&\qquad \left.+ \varGamma_{2,2} \left[^1_1\right]~(\vartheta_4^8-\vartheta_2^8)~\bar\vartheta_4^4\,\bar\vartheta_2^4\,\left( (\hat{\bar E}_2+\bar\vartheta_4^4)\,\bar\vartheta_2^4\,\bar\vartheta_4^4-8\bar\eta^{12}\right) \right] \,,
\end{split}
\end{equation}
where $\hat E_2$ is the weight-two quasi holomorphic Eisenstein series\footnote{Whenever the characteristics of the theta constants equal $0, 1/2$ we employ the light notation in terms of the $\vartheta_\alpha$'s.}.
Notice that the second and third lines can be obtained from the first one upon acting with the ${\rm SL} (2;\mathbb{Z})$ generators $S$ and $TS$, as demanded by modularity.

The remaining contributions can be organised as
\begin{equation}
\begin{split}
\sum_{H,G =0}^1 \sum_{h,g = 0\atop (h,g)\neq (0,0)}^{N-1} \varDelta_\mathcal{G} \left[ ^{H\,, \ h}_{G\,, \ g}\right] &= \sum_{h,g = 0\atop (h,g)\neq (0,0)}^{N-1} \left(
\varDelta_\mathcal{G} \left[ ^{0\,, \ h}_{0\,, \ g}\right] + 
\varDelta_\mathcal{G} \left[ ^{0\,, \ h}_{1\,, \ g}\right] +
\varDelta_\mathcal{G} \left[ ^{1\,, \ h}_{0\,, \ g}\right] +
\varDelta_\mathcal{G} \left[ ^{1\,, \ h}_{1\,, \ g}\right] 
\right)
\\
&\equiv \varDelta_\mathcal{G}^{(u_+)} + \varDelta_\mathcal{G}^{(u_-)} + \varDelta_\mathcal{G}^{(t_+)} + \varDelta_\mathcal{G}^{(t_-)} \,,
\end{split}
\end{equation}
according to the sectors of the freely-acting orbifold. The first contribution $\varDelta_\mathcal{G}^{(u_+)}$, corresponding to $(H,G)=(0,0)$, computes the gauge thresholds of the $\mathcal{N}=2$ heterotic string on the orbifold limit of ${\rm K3}$. It is thus expected to be BPS saturated and the difference $\varDelta_\mathcal{G}^{(u_+)} - \varDelta_{\mathcal{G}'}^{(u_+)}$ to be universal and to depend only on the moduli of the $T^2$ torus\footnote{The difference of gauge thresholds is indeed universal for the $T^4 /\mathbb{Z}_N$ orbifolds, though in more general constructions they may exhibit a non-universal structure \cite{Mayr:1993mq,KiritsisEN}.} \cite{Dixon:1990pc}. The remaining terms, connected among each other by $S$ and $TS$ modular transformations, are inherently  
non-BPS since the freely-acting orbifold acts non-trivially and breaks supersymmetry. This is reflected by the fact that the modular integral now involves genuinly non-holomorphic contributions.

For concreteness, we shall present explicitly the various contributions in the case $N=2$, where one finds
\begin{equation}
\varDelta_{{\rm SO} (16)}^{(u_+)} =  -\frac{1}{24}\,\varGamma_{2,2} \left[^0_0\right] \,\frac{\hat{\bar E}_2\,\bar E_4\,\bar E_6-\bar E_6^2}{\bar\eta^{24}}\,,
\label{uplus16}
\end{equation}
\begin{equation}
\begin{split}
\varDelta_{{\rm SO} (16)}^{(u_-)} &= -\frac{1}{48}\,\varGamma_{2,2} \left[ ^0_1\right] \,
\frac{\bar\vartheta_3^4\,\bar\vartheta_4^4(\bar\vartheta_3^4+\bar\vartheta_4^4) \left[ (\hat{\bar E}_2-\bar\vartheta_3^4)\,\bar\vartheta_3^4\,\bar\vartheta_4^4+8\bar\eta^{12}\right]}{\bar \eta^{24}}
\\
&\qquad -\frac{1}{72}\,\varGamma_{2,2} \left[ ^0_1\right] \frac{\vartheta_2^4\,(\vartheta_3^8-\vartheta_4^8)}{\eta^{12}}
\,\frac{(\hat{\bar E}_2-\bar\vartheta_3^4)\,\bar\vartheta_3^4\,\bar\vartheta_4^4+8\bar\eta^{12}}{\bar\eta^{12}} \,,
\end{split}\label{uminus16}
\end{equation}
and 
\begin{equation}
\varDelta_{{\rm SO} (12)}^{(u_+)} = -\frac{1}{24}\,\varGamma_{2,2} \left[^0_0\right] \,\frac{\hat{\bar E}_2\,\bar E_4\,\bar E_6-\bar E_4^3}{\bar\eta^{24}}\,,
\label{uplus12}
\end{equation}
\begin{equation}
\begin{split}
\varDelta_{{\rm SO} (12)}^{(u_-)} &=
-\frac{1}{48}\, \varGamma_{2,2}\left[ ^0_1 \right] \, \frac{\bar\vartheta_3^8\,\bar\vartheta_4^8\left[\hat{\bar E}_2\,(\bar\vartheta_3^4+\bar\vartheta_4^4)+\bar\vartheta_2^8-2\bar\vartheta_3^4\,\bar\vartheta_4^4\right] }{\bar\eta^{24}}
\\
&-\frac{1}{72}\, \varGamma_{2,2}\left[ ^0_1 \right] \, \left(
\frac{\vartheta_2^4\,(\vartheta_3^8-\vartheta_4^8)}{\eta^{12}}\,\frac{\hat{\bar E}_2\,\bar\vartheta_3^4\,\bar\vartheta_4^4}{\bar\eta^{12}}  +\frac{\vartheta_2^4\,\vartheta_4^4\,|\vartheta_2^4-\vartheta_4^4|^2-\vartheta_2^4\,\vartheta_3^4\,|\vartheta_2^4+\vartheta_3^4|^2}{\eta^{12}\,\bar\eta^{12}}\,\bar\vartheta_3^4\,\bar\vartheta_4^4 \right) \,.
\end{split}\label{uminus12}
\end{equation}
In these expressions $E_4$  ($E_6$) is the weight-four (-six) holomorphic Eisenstein series. 
Again, the remaining terms can be computed by the action of the generators $S$ and $TS$ of ${\rm SL} (2;\mathbb{Z})$ on the corresponding $\varDelta^{(u_-)} $ contributions.

As anticipated, eqs. \eqref{uplus16} and \eqref{uplus12} compute the thresholds to the $\mathcal{N}=2$ supersymmetric ${\rm E}_8$ and ${\rm E}_7$ gauge factors. 
The eqs. \eqref{uminus16} and \eqref{uminus12} involve contributions from BPS states whose masses are now deformed by the free action of the $\mathbb{Z}_2^\prime$ orbifold.

The BPS contributions to these amplitudes can  be integrated over the ${\rm SL} (2;\mathbb{Z})$ fundamental domain $\mathcal{F}$ or, after partial unfolding, over the fundamental domain $\mathcal{F}_0 [2]$ of the $\varGamma_0 (2)$ congruence subgroup, following the procedure developed in \cite{Angelantonj:2011br, Angelantonj:2012gw, Angelantonj:2013eja}. The non-BPS contributions can be shown to be exponentially suppressed \cite{AFT} in the large $T^2$ volume limit, and are thus negligible at low-energies. We shall not indulge here in the full computation of the thresholds, but rather focus on their difference. One finds
\begin{equation}
\varDelta_{{\rm SO} (16)}^{(u_+)} - \varDelta_{{\rm SO} (12)}^{(u_+)} = -72 \, \varGamma_{2,2} \left[^0_0\right] \,,
\end{equation}
that reproduces the result of \cite{Dixon:1990pc}, and
\begin{equation}
\varDelta_{{\rm SO} (16)}^{(u_-)} - \varDelta_{{\rm SO} (12)}^{(u_-)} = - \tfrac{1}{3} \, \varGamma_{2,2} \left[ ^0_1\right] \, \left(\frac{\vartheta_2^{12}}{\eta^{12}}  -8 \right)\,.
\label{holcont}
\end{equation}
Surprisingly, the non-holomorphic contributions to the thresholds cancel when taking   their difference, and reduce to a purely  holomorphic BPS-like  term. As we shall show, the difference of gauge thresholds  exhibits a remarkable universal behaviour, independently of the details of the $T^4/\mathbb{Z}_N$ orbifold. Indeed, the non-holomorphic contribution to the difference of thresholds reads
\begin{equation}
\begin{split}
& -\frac{\vartheta_2^8\,|\vartheta_3^4+\vartheta_4^4|^2\,\bar\vartheta_3^4\,\bar\vartheta_4^4}{\eta^{12}\,\bar\eta^{12}}  -\frac{\vartheta_2^4\,\vartheta_4^4|\vartheta_2^4-\vartheta_4^4|^2\,\bar\vartheta_3^4\,\bar\vartheta_4^4}{\eta^{12}\,\bar\eta^{12}}+
	\frac{\vartheta_2^4\,\vartheta_3^4\,|\vartheta_2^4+\vartheta_3^4|^2\,\bar\vartheta_3^4\,\bar\vartheta_4^4}{\eta^{12}\,\bar\eta^{12}} 
\\
&\qquad \qquad = 12\,(O_8^2\,V_8+3V_8^3)\, (\bar O_8^2\,\bar V_8-\bar V_8^3)\,,
\end{split}
\label{nonhol}
\end{equation}
where in the right-hand side we have introduced the ${\rm SO} (8)$ characters. Although, this term looks completely non-holomorphic it actually possesses a BPS-like structure due to a remarkable MSDS identity \cite{Florakis:2009sm, Florakis:2010ty}
\begin{equation}
\bar O_8^2\,\bar V_8-\bar V_8^3 = 8\,,
\end{equation}
which reflects a hidden MSDS spectral flow in the bosonic sector of the global $\mathcal{N} = (2,2)$ superconformal symmetry on the world-sheet \cite{Faraggi:2011aw,AFT}. 
As a result, eq. \eqref{nonhol} reduces to the purely holomorphic contribution \eqref{holcont}. 

To evaluate the integrals, we first notice that the combination $\vartheta_2^{12}/\eta^{12}$ corresponds to an automorphic function of the Hecke congruence subgroup $\varGamma_0 (2)$. Moreover, it is regular at the cusp at $\tau = i \infty$ while it has a simple pole at the cusp $\tau =0$\footnote{We remind here that the compactification of the fundamental domain $\cF_0 [2]$ of $\varGamma_0 (2)$ requires adding two points, {\it i.e.} the two cusps, $\tau = i \infty$ and $\tau = 0$. See, for instance, \cite{Angelantonj:2013eja}.}. This is sufficient to identify \cite{Angelantonj:2013eja}
\begin{equation}
\frac{\vartheta_2^{12}}{\eta^{12}} = \cF_0 (1,1,0)  - 16 = \hat \jmath_2 (\tau ) -24 \,,
\end{equation}
where $\cF_0 (1,1,0)$ is the meromorphic weight-zero Niebur-Poincar\'e series attached to the cusp at $\tau =0$ of $\varGamma_0 (2)$, and $\hat\jmath_2 (\tau) $ is the Fricke transform  \cite{Angelantonj:2013eja} of the $\varGamma_0 (2)$ Hauptmodul
\begin{equation}
j_2 (\tau ) = \frac{\eta^{24} (\tau ) }{\eta^{24} (2\tau )}  + 24\,.
\end{equation}

The modular integrals can be straightforwardly computed using the results of\footnote{The first integral was actually originally computed in \cite{Dixon:1990pc} by unfolding the fundamental domain against the Narain lattice.}  \cite{  Angelantonj:2011br, Angelantonj:2012gw ,Angelantonj:2013eja,  AFPnew} to yield
\begin{equation}
{\rm R.N.}\int_{\mathcal F}d\mu\,\varGamma_{2,2}(T,U) = -\log T_2 U_2|\eta(T)\,\eta(U)|^4\,,
\end{equation}
\begin{equation}
{\rm R.N.}\int_{\mathcal F_0[2]}d\mu\,\varGamma_{2,2}[^0_1](T,U)= -\log T_2 U_2 |\vartheta_4(T)\,\vartheta_2(U)|^4\,,
\end{equation}
and
\begin{equation}
{\rm R.N.}\int_{\mathcal F_0[2]}d\mu\, \varGamma_{2,2}[^0_1](T,U)\,\frac{\theta_2^{12}(\tau)}{\eta^{12}(\tau)}=-2\log|j_2(T/2)- j_2(U)|^4 \,.
\end{equation}
Combining the various contributions, one finds
\begin{equation}
\begin{split}
\varDelta_{{\rm SO} (16)} - \varDelta_{{\rm SO} (12)} &= 72\,\log \left[ T_2 U_2|\eta(T)\,\eta(U)|^4\right]-\tfrac{8}{3}\,\log \left[ T_2 U_2|\vartheta_4(T)\,\vartheta_2(U)|^4 \right]
\\
&\qquad +\tfrac{2}{3}\,\log | j_2(T/2)-j_2(U)|^4 \,.
\end{split}\label{thresh}
\end{equation}
Again, the various terms have a clear physical interpretation. The first line generalises the celebrated result of \cite{Dixon:1990pc}.  The presence of second term is ascribed to the modified Kaluza-Klein masses of BPS states, that are indeed affected by the free action of the orbifold. In fact, since the $\mathbb{Z}_2^\prime$ orbifold corresponds to a spontaneous breaking of supersymmetry, the model in \eqref{partition} contains precisely the same  excitations as the  ${\rm E_8}\times {\rm E}_8$ heterotic string on $T^2\times T^4 /\mathbb{Z}_N$, whose masses are continuously deformed by the scale of supersymmetry breaking. As a result, the duality group is broken down to the subgroup $\varGamma^0 (2)_T \times \varGamma_0 (2)_U$ of 
${\rm SL} (2;\mathbb{Z})_T \times {\rm SL}  (2;\mathbb{Z} )_U$.

While the first two contributions are regular at any point in the classical moduli space, the term in the second line, particular to this vacuum with broken supersymmetry, possesses logarithmic singularities at the locus $T/2 = U$ and its $\varGamma_0 (2)$ images. The origin of these singularities is ascribed to massive charged BPS-like states that become massless at special points in moduli space.  To manifest their origin in the perturbative spectrum, it is convenient to express their contribution to \eqref{partition} in terms of the ${\rm SO} (2n)$ characters
\begin{equation}
\tfrac{1}{2}\, (O_4\,O_4\times \bar V_{12}\,\bar O_4 \,\bar V_{16}) \, \left( \varGamma_{2,2}[^1_0] + \varGamma_{2,2} [^1_1] \right) \,.
\end{equation}
These states include the left-moving NS vacuum and its stringy excitations, while the right-moving sector is massless and belongs to  the bi-fundamental representation $(16,12)$ of the ${\rm SO} (16) \times {\rm SO} (12)$ gauge group. They always carry non-trivial momentum and winding quantum numbers, and the lightest states  have mass
\begin{equation}
m^2_{O_4 O_4} = \frac{|T/2-U|^2}{T_2 U_2}\,.
\end{equation}
Indeed, these states become massless at the point $T/2=U$, where $p_{\rm R}^2=0$, and are responsible for the logarithmic divergence in \eqref{thresh}. 

Notice that the fact that extra massless states emerge from the $\mathbb{Z}_2^\prime$ twisted sector  is compatible with the fact that the term $\vartheta_2^{12}/\eta^{12}$, originating from  the un-twisted sector, has a pole at the cusp $\tau = 0$ but is regular at $\tau =i \infty$. In fact, the two cusps are related by an $S$ modular transformation that also relates the untwisted and twisted sectors. As a result, the  singularity of the $u_-$ sector at $\tau =0$ is to be understood as the map under $S$ of the physical  infra-red singularity of the twisted sector. 

One can compute the gauge thresholds also in the case of the other singular limits of K3, namely $N=3,4,6$. Although the result of the thresholds depends on the particular value of $N$, their difference exhibits a remarkable {\em universal} behaviour. In fact, one finds
\begin{equation}
\begin{split}
\varDelta_{{\rm SO} (16)} - \varDelta_{{\rm SO} (12) }  &= \alpha \,\log \left[ T_2 U_2|\eta(T)\,\eta(U)|^4\right]+\beta\,\log \left[ T_2 U_2|\vartheta_4(T)\,\vartheta_2(U)|^4 \right]
\\
&\qquad +\gamma\,\log|j_2(T/2)-j_2(U)|^4  \,,
\end{split}\label{universal}
\end{equation}
with $(\alpha , \beta, \gamma) = (72 , -\tfrac{8}{3} , \tfrac{2}{3})$ for the $\mathbb{Z}_2$ and $\mathbb{Z}_3$ orbifolds, $(\alpha , \beta, \gamma) =\tfrac{5}{8} (72 , -\tfrac{8}{3} , \tfrac{16}{15})$ for the $\mathbb{Z}_4$ orbifold and $(\alpha , \beta, \gamma) =\tfrac{35}{144} (72 , -\tfrac{8}{3} , \tfrac{2}{3})$ for the $\mathbb{Z}_6$ orbifold.

This universality structure is a direct consequence of the universal behaviour of the $\mathcal{N}=2$ thresholds \cite{Dixon:1990pc,Kiritsis:1996dn}, which is preserved by the free action of the $\mathbb{Z}_2^\prime$.

\section{Decompactification limits}

It is instructive to study eq. \eqref{universal} in the decompactification limits. For convenience, we shall assume a squared $T^2$ with 
\begin{equation}
T= i\,R_1 R_2 \qquad {\rm and} \qquad U=i\,\frac{R_2}{R_1} \,,
\end{equation}
so that the masses of the two gravitini and of the $O_4 O_4$ states read
\begin{equation}
m_{3/2}^2 = \frac{1}{R^2_1}\qquad {\rm and}\qquad m_{O_4 O_4}^2 = \frac{1}{4} \left( R_1 - \frac{2}{R_1}\right)^2\,.
\end{equation}

In the $R_1  \to \infty$ limit, $\mathcal{N} =2$ supersymmetry is recovered, and the leading behaviour of eq. \eqref{universal} 
\begin{equation}
\lim_{R_1\rightarrow\infty}\left[\varDelta_{{\rm SO} (16)} - \varDelta_{{\rm SO} (12)}\right]=\frac{\pi\alpha}{3}\,R_1\left(R_2+\frac{1}{R_2}\right) + \ldots
\end{equation}
grows linearly with the $T^2$ volume. This is expected from scaling arguments, since in six dimensions the gauge coupling has length dimension $-1$. The term proportional to $\beta $ in \eqref{universal} only grows logarithmically with $R_1$ as a result of supersymmetry enhancement since, charged states lighter than the supersymmetry-breaking scale are effectively BPS-like and thus contribute logarithmically to the  difference of threshold corrections, whereas the infinite tower of charged states heavier than $m_{3/2}$ have an effective $\mathcal{N}=4$ supersymmetry and, thus, do  not contribute. Finally, the term proportional to $\gamma$ is exponentially suppressed because the lightest charged states $O_4O_4$ have mass  $m_{O_4 O_4 } \gg m_{3/2}$ and effectively decouple.

In the $R_2 \to \infty$ limit,  the  leading behaviour of \eqref{universal} is
\begin{equation}
\begin{split}
\lim_{R_2\rightarrow\infty}\left[\varDelta_{{\rm SO} (16)} - \varDelta_{{\rm SO} (12)} \right] &=\frac{\pi\alpha}{3}\,R_2\left(R_1+\frac{1}{R_1}\right)+\frac{\pi\beta R_2}{R_1} 
\\
&\qquad +2\pi\gamma\, R_2\, \left(R_1-\frac{2}{R_1}-\left|R_1-\frac{2}{R_1}\right|\right)+\ldots\,.
\end{split}
\end{equation}
As expected, the term proportional to $\alpha$ is again linearly divergent with the $T^2$ volume. The term proportional to $\beta$ now scales as $R_2/R_1$,  and consistently vanishes as $m_{3/2} \to 0$. The term proportional to $\gamma$ depends on the scale of supersymmetry breaking. When $R_1>\sqrt{2}$ it is  exponentially suppressed because $m_{3/2} < m_{O_4 O_4}$, whereas when $R_1< \sqrt{2}$ it scales as $R_2(2/R_1-R_1)$. This is a consequence of the fact that, in the $R_1\to 0$ limit, supersymmetry is explicitly broken, and this term grows with the volume $R_2 \tilde{R}_1 \sim R_2/R_1 $ of the T-dual torus.

Notice that in the $R_1 \to 0 $ limit, the freely-acting orbifold degenerates into an explicit breaking of supersymmetry. This implies that the universal behaviour \eqref{universal} should hold also in the case when the ten-dimensional ${\rm O} (16) \times {\rm O} (16)$ theory of \cite{AlvarezGaume:1986jb, Dixon:1986iz} is compactified on  $T^2\times {\rm K}3$. As a result, a similar universal behaviour of the threshold differences is expected to arise also when $T^2\times {\rm K}3$ is replaced by a generic Calabi-Yau manifold. It would be interesting to investigate whether eq. \eqref{universal} also holds when the ten-dimensional heterotic string, whether supersymmetric or not, is compactified on a manifold that does not preserve any supersymmetry.

\subsection*{\bf Acknowledgments}
We are grateful to B. Pioline for fruitful discussions. 
M.T. would like to thank the Department of Physics, the University of Auckland, where part of this work has
been performed, for its kind hospitality. 
The work of C.A. has been supported in part by the European ERC Advanced Grant no. 226455 ``Supersymmetry, Quantum Gravity and Gauge Fields'' (SUPERFIELDS)
and in part by the Compagnia di San Paolo contract ``Modern Application in String Theory'' (MAST) TO-Call3-2012-0088.
The work of M.T.  has been supported in part by an  Australian Research Council  grant DP120101340. M.T. would also like to acknowledge grant   31/89 of the Rustaveli National Science Foundation.


\bibliographystyle{unsrt}

\end{document}